\documentclass[revsymb,10pt,prb]{revtex4}%
\usepackage{graphicx}
\usepackage{amsmath}
\usepackage{amsfonts}
\usepackage{amssymb}%
\providecommand{\U}[1]{\protect\rule{.1in}{.1in}}
\newcommand{\be}{\begin{equation}}
\newcommand{\en}{\end{equation}}

\begin{document}
\title{Bose-Einstein condensation of semi-hard bosons in S=1 dimerized organic
compound F$_2$PNNNO}
\author{I.G. Bostrem, V.E. Sinitsyn, A.S. Ovchinnikov}

\address{Department of Physics, Ural State University, 620083, Ekaterinburg, Russia}

\author{Y. Hosokoshi}
\address{Department  of Physical Science, Osaka Prefecture University, Osaka, Japan}
\author{K. Inoue}
\address{Department of Chemistry, Hiroshima University, Hiroshima, Japan}
\date{\today }

\begin{abstract}
An analysis of the energy spectrum and the magnetization curve of two-dimensional organic antiferromagnet F$_2$PNNNO with a spin-one dimerized structure shows that a behavior of the compound in an external magnetic field can be explained within a lattice boson model with an extended Pauli's exclusion principle, i.e. no more than two bosons per a dimer. The unusual magnetization curve observed experimentally in the compound reflects a sequence of phase transitions intrinsic for a lattice boson system with  strong on-site and inter-site repulsions due to a tuning of magnon density by the applied magnetic field.
\end{abstract}

\pacs{Valid PACS appear here}
\maketitle

\section{Introduction}

A possibility to study the Bose-Einstein condensation (BEC) with
low-dimensional magnetic materials predicted theoretically twenty years ago%
\cite{Affleck} gave rise to intense experimental studies in the field. The
analogy between the spins and the bosons becomes evident for
antiferromagnets where spins form dimers with a spin-singlet ground state.%
\cite{Giamarchi} Originally, main attention was focused on spin-1/2 systems
where excitations inside each dimer (triplons) are regarded as bosons with
hard-core repulsion, i.e. no more than one boson present on a single dimer.
The analogy enables to treat the spin systems as that of interacting bosons
whose ground state is determined by the balance between the kinetic energy
and the repulsive interactions.\cite{Rice} If the repulsion dominates the
bosons will form a superlattice and a finite energy cost is needed to create
an additional particle. This exhibits itself as a jump in chemical potential
versus boson number, in the spin language, as a plateau in magnetization
curve versus magnetic field at rational fraction of saturated magnetization.

The field induced condensation of magnons has been experimentally observed
in coupled quantum ($s=1/2$) dimer systems based on Cu$^{2+}$ ions such as
TlCuCl$_3$ and BaCuSi$_2$O$_6$ \cite{Nikuni,Ruegg,Sebastian} and the
compound Ba$_2$Cr$_2$O$_8$ \cite{Kofu} which are adequately described by the
BEC theory.

Recently, magnetic weakly coupled dimer system Ba$_3$Mn$_2$O$_8$ with $S=1$
moments attracted a lot of attention.\cite{Andraka,Stone} The field behavior
of magnetization in the system of antiferromagnetically weakly coupled $S=1$
dimers can be described as BEC of magnons by mapping the spin-1 system into
a gas of semi-hard-core bosons.\cite{Batista} On an example of simple
two-dimensional (2D) $S=1$ isotropic Heisenberg model with a dimerized structure
and frustrating interactions it was suggested an emergence of the spin
supersolid state (a long-range mixing of superfluid and charge ordered phases)
induced by a magnetic field.\cite{Sengupta}

The organic compound F$_2$PNNNO is a supplementary example of spin-one dimer based
magnetic insulator. This is 2D Heisenberg system with a singlet ground state,
in which $S=1$ dimers interact antiferromagnetically.\cite{Hosokoshi,Tsujii}
The lattice of the system is  equivalent to the  honeycomb one
(Fig. \ref{chessboard}). The field magnetization process shows a two-step saturation
behavior that is a rare example of observation of a plateau in a
two-dimensional system. The intermediate plateau corresponds to the half
value of saturation magnetization. The consistent calculation of
susceptibility and magnetization for the  finite-size cluster with imposed periodic conditions  yields the
following estimations of antiferromagnetic exchange couplings $2J_0=67.5$
K, $2J_1=7.5$ K, i.e. the system can be regarded as a real 2D dimerized
spin-one system.

Apparently, the quantum antiferromagnet F$_2$PNNNO offers an opportunity to
verify a relevance of semi-hard core boson model for description of the
dimerized system. In the paper we perform a diagonalization of finite
cluster of $N=18$ sites, calculate the magnetization and demonstrate that
these results can be easily understood within the semi-hard boson model with strong
on-site and inter-site repulsions. The diagonalization procedure used by us accounts the spin rotational symmetry.\cite
{Boyarchenkov,Sinitsyn} The implementation of non-Abelian $SU(2)$ spin
symmetry is based on an elimination of quantum numbers via the Wigner-Eckart
theorem. The advantage of the approach is that the cluster spin states are decomposed into  different sectors of the total cluster    spin. In addition, one can independently handle each of the target spin state.

The paper is organized as follows. The model and the diagonalization algorithm are given in Sec.II. The truncation procedure is discussed  in Sec.III. In Sec.IV we report numerical  cluster calculations of the spectrum and the magnetization curve. The analogy with the lattice  boson model is performed in Sec.V. Main results  are recapitulated in the Conclusion part.

\section{The model}

\begin{figure}[ptb]
\begin{center}
\includegraphics[width=85mm]{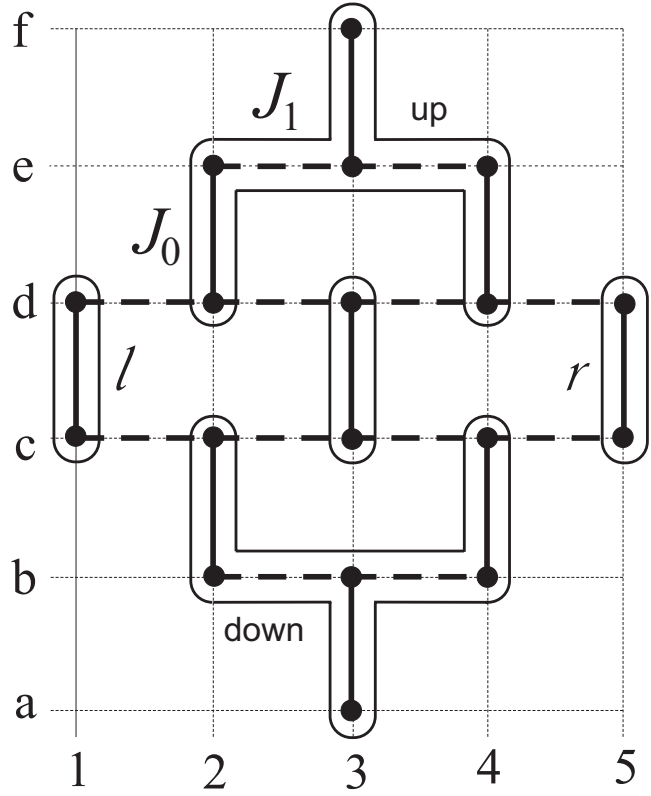}
\end{center}
\caption{The 18-site cluster used in numerical calculations. The environment of the central dimer consists of two "fork"-like parts (up and down), and the left ($l$) and right ($r$) dimers. The intra-dimer $J_0$ and inter-dimer $J_1$ interactions   are shown by solid and dotted lines, respectively.}
\label{chessboard}
\end{figure}

The Hamiltonian of weakly interacting spin-one dimers on a 2D lattice
depicted in Fig. \ref{chessboard} is given by 
\begin{equation}
H_S=J_0\sum\limits_{i}\vec{S}_{i1 }\vec{S}_{i2
}+J_1\sum\limits_{\left\langle i\alpha,  j {\bar \alpha} \right\rangle } \vec{S}_{i\alpha
}\vec{S}_{j\bar{\alpha}},  \label{Hspin}
\end{equation}
where $J_0$ is the coupling inside the $i$-th dimer, $J_1$ is the strength of the
exchange interaction between the dimers located on the bonds $\left\langle
i,j\right\rangle $. The indices $\alpha$, $\bar \alpha$ mark $S=1$ spins that enter
into the interacting dimers, namely, $\bar{\alpha}$ = 1,2 provided $\alpha$ = 2,1, respectively. The both types of the  interactions are antiferromagnetic $J_{0,1}>0$, and the regime of weakly interacting dimers,  $\left|
J_0\right| \gg \left| J_1\right| $, is considered. The Heisenberg model has been previously suggested  to explain some thermodynamical properties of F$_2$PNNNO.\cite{Hosokoshi} Numerical calculations  based on the Hamiltonian (\ref{Hspin}) via exact diagonalization of small clusters and their comparison with experimental data prove its relevance for the ratio   $\left| J_1/J_0 \right| \ll 1$. The dimerization caused by the anisotropy of interactions on a lattice is somewhat analogous to a situation in two-leg spin-1 antiferromagnetic ladders in a strong antiferromagnetic rung-coupling regime, when the ladder ground state is well approximated by the tensor product of singlet rung-dimers.\cite{Sato2005}

To get the energy spectrum the finite-size clusters composed of $N=10$ and $N=18$ sites are selected. In a choice of the cluster care should be taken to ensure that the lattice point group symmetry is hold. Since intra-dimer interactions are the strongest, the cluster should consist of intact dimers.  To mark  sites inside the cluster the chessboard-like notations will be used, i.e. site positions along the $x$ axis are marked by numbers whereas positions along the y axis are denoted by Latin letters.

To find eigenfunctions of the cluster that inherit the total cluster spin as
a quantum number we should develop a consecutive procedure of addition of
spin moments. It is convenient to break the cluster in several parts.
Following the strategy of a cluster building used in Ref.\cite{Sinitsyn},
one should identify the central dimer (center) and its environment. The
center is composed of the $c3$ and $d3$ sites whereas another sites are
embodied into the environment.

The Hamiltonian of the central dimer has the form $H_c=J_0\vec{S}_{c3}\vec{S}%
_{d3},$ whereas the interaction between the center and its environment is
given by 
\begin{equation}
V_{\text{ce}}=J_1\vec{S}_{c3}\left( \vec{S}_{c2}+\vec{S}_{c4}\right) +J_1\vec{S}%
_{d3}\left( \vec{S}_{d2}+\vec{S}_{d4}\right) .  \label{2_2}
\end{equation}
The environment consists of four parts, namely of two dimers, left ($l$) and
right ($r$) ones, with the Hamiltonians 
\begin{equation}
H_l=J_0\vec{S}_{c1}\vec{S}_{d1},\;\text{and}\;H_r=J_0\vec{S}_{c5}\vec{S}%
_{d5},  \label{2_3}
\end{equation}
respectively, as well as two fork-like parts, i.e. the down  and upper ones, with the corresponding Hamiltonians 
\begin{equation}
H_{\text{down}}=J_0\left( \vec{S}_{b2}\vec{S}_{c2}+\vec{S}_{a3}\vec{S}_{b3}+\vec{S}_{b4}%
\vec{S}_{c4}\right) +J_1\vec{S}_{b3}\left( \vec{S}_{b2}+\vec{S}_{b4}\right) ,
\label{2_5}
\end{equation}
\begin{equation}
H_{\text{up}}=J_0\left( \vec{S}_{d2}\vec{S}_{e2}+\vec{S}_{e3}\vec{S}_{f3}+\vec{S}_{d4}%
\vec{S}_{e4}\right) +J_1\vec{S}_{e3}\left( \vec{S}_{e2}+\vec{S}_{e4}\right) .
\label{2_6}
\end{equation}
The interaction between the left/right dimers and the fork-like parts is
presented as  
\begin{equation}
V_{\text{env}}=J_1\left( \vec{S}_{c2}\vec{S}_{c1}+\vec{S}_{d2}\vec{S}_{d1}+\vec{%
S}_{c4}\vec{S}_{c5}+\vec{S}_{d4}\vec{S}_{d5}\right) .  \label{2_7}
\end{equation}
The Hamiltonian of the entire cluster gathers all the above terms 
\begin{equation}
H=H_c+V_{\text{ce}}+\left\{ H_l+H_r+H_{\text{down}}+H_{\text{up}}+V_{\text{env}}\right\} .  
\label{2_8}
\end{equation}

There are three states of the dimer, which is the elementary block of the cluster,
 with the total spin $S_{\text{dm}}=0$ (singlet), $S_{\text{dm}}=1$
(triplet)$,$ and $S_{\text{dm}}=2$ (quintiplet). The energies of the states are $%
E_0=-2J_0$, $E_1=-J_0$, $E_2=J_0$, respectively, and the eigenstates are
obtained via the common rule of addition of moments 
\begin{equation}
\left| 11;S_{\text{dm}}M_{\text{dm}}\right\rangle \equiv
\left| S_{\text{dm}}M_{\text{dm}}\right\rangle
=\sum\limits_{\sigma _1\sigma _2}\left[ 
\begin{array}{ccc}
1 & 1 & S_{\text{dm}} \\ 
\sigma _1 & \sigma _2 & M_{\text{dm}}
\end{array}
\right] \left| 1\sigma _1\right\rangle \left| 1\sigma _2\right\rangle,
\label{2_9}
\end{equation}
where $\left[ \ldots \right] $ is the Clebsch-Gordan coefficient. To
increase the cluster size the reduced matrix elements (RME) of the spin
operators $S(1)$ and $S(2)$, that constitute the dimer, calculated within the basis (%
\ref{2_9}) are needed 
\begin{equation}
\left\langle S_{\text{dm}} \left\| S(1) \right\| S_{\text{dm}}^{^{\prime }}\right\rangle =\left(
-1\right) ^{1+S_{\text{dm}}^{\prime }}\sqrt{(2S_{\text{dm}}+1)(2S_{\text{dm}}^{\prime }+1)}\left\{ 
\begin{array}{ccc}
S_{\text{dm}} & 1 & S_{\text{dm}}^{\prime } \\ 
1 & 1 & 1
\end{array}
\right\} \left\langle 1\left\| S\right\| 1\right\rangle ,  \label{2_10}
\end{equation}
\begin{equation}
\left\langle S_{dm}\left\| S(2)\right\| S_{dm}^{^{\prime }}\right\rangle =\left(
-1\right) ^{1+S_{dm}}\sqrt{(2S_{dm}+1)(2S_{dm}^{\prime }+1)}\left\{ 
\begin{array}{ccc}
S_{dm} & 1 & S_{dm}^{\prime } \\ 
1 & 1 & 1
\end{array}
\right\} \left\langle 1 \left\| S \right\| 1\right\rangle ,  \label{2_11}
\end{equation}
where $\left\{ \ldots \right\} $ is the $6j$-symbol of the rotation group, and the reduced matrix element
$\left\langle 1\left\| S \right\| 1\right\rangle = \sqrt{6}$.

The fork-like part includes three interacting dimers. It is convenient to
build the basis of the fragment according to the scheme $(2+4)+3$ of the
moment addition, i.e. a combining of the ''prong'' dimer functions is
followed by adding of the ''handle'' function. As a result, the basic
functions with the total spin $S_{\text{down}}$ of the down fork-like part has the form
\begin{equation}
\left| \left( S_2 S_4\right) S_{24},S_3;S_{\text{down}}M_{\text{down}}\right\rangle
=\sum\limits_{M_2 M_3 M_4 M_{24}}\left[ 
\begin{array}{ccc}
S_2 & S_4 & S_{24} \\ 
M_2 & M_4 & M_{24}
\end{array}
\right] \left[ 
\begin{array}{ccc}
S_{24} & S_3 & S_{\text{down}} \\ 
M_{24} & M_3 & M_{\text{down}}
\end{array}
\right] \left| S_2M_2\right\rangle \left| S_3M_3\right\rangle \left|
S_4M_4\right\rangle ,  \label{2_12}
\end{equation}
where $S_2$, $S_3$ and $S_4$ are the spins of the dimers composed of the $b2$
and $c2$ sites, etc. Within the basis the Hamiltonian (\ref{2_5}) is
presented by the block diagonal $141\times 141$ matrix. The blocks are
marked by the total spin $S_{\text{down}}=0,1,\ldots ,6$ values. A diagonalization of
the $H_{\text{down}}$ matrix yields the spectrum $E_{i_{\text{down}} S_{\text{down}}}$ and the eigenfunctions 
\[
\left| i_{\text{down}} S_{\text{down}} M_{\text{down}} \right\rangle =\sum\limits_{S_2S_3S_4S_{24}}\alpha _{\left(
S_2S_4\right) S_{24},S_3}^{i_{\text{down}} S_{\text{down}}}\left| \left( S_2S_4\right)
S_{24},S_3;S_{\text{down}} M_{\text{down}}  \right\rangle , 
\]
where the index \thinspace $i_{\text{down}}$ distinguishes basic functions with the same
total $S_{\text{down}}$ spin. The results for the upper fork-like part can be obtained
analogously provided the site $c4$ is substituted for $d2$, and $c2$ is
changed by $d4$ etc. The assembly of the cluster part is completed by calculations of the reduced
matrix elements [see Eq.(\ref{A_1}) in Appendix].

As the next step, we construct the spin functions of the non-interacting parts,
i.e. of the left and the right dimers
\begin{equation}
\left| S_lS_r;S_{lr}M_{lr}\right\rangle =\sum\limits_{M_lM_r}\left[ 
\begin{array}{ccc}
S_l & S_r & S_{lr} \\ 
M_l & M_r & M_{lr}
\end{array}
\right] \left| S_lM_l\right\rangle \left| S_rM_r\right\rangle ,  \label{2_13}
\end{equation}
where $S_{lr}=0,1\ldots ,4$, and the upper and down fork-like parts 
\begin{equation}
\left| i_{\text{up}} S_{\text{up}} i_{\text{down}} S_{\text{down}};S_{\text{ud}}M_{\text{ud}} \right\rangle =\sum\limits_{M_{\text{up}} M_{\text{down}}}\left[ 
\begin{array}{ccc}
S_{\text{up}} & S_{\text{down}} & S_{\text{ud}} \\ 
M_{\text{up}} & M_{\text{down}} & M_{\text{ud}}
\end{array}
\right] \left| i_{\text{up}} S_{\text{up}} M_{\text{up}} \right\rangle \left| i_{\text{down}} S_{\text{down}} M_{\text{down}} \right\rangle ,
\label{2_14}
\end{equation}
where $S_{\text{ud}}=0,1\ldots ,12$, and add them together to build the basis of
the environment of the central dimer
\[
\left| \left( i_{\text{up}} S_{\text{up}} i_{\text{down}} S_{\text{down}} \right) S_{\text{ud}},\left( S_lS_r\right) S_{lr};S_{%
\text{env}}M_{\text{env}}\right\rangle 
\]
\begin{equation}
=\sum\limits_{M_{\text{ud}} M_{lr}}\left[ 
\begin{array}{ccc}
S_{\text{ud}} & S_{lr} & S_{\text{env}} \\ 
M_{\text{ud}} & M_{lr} & M_{\text{env}}
\end{array}
\right] \left| i_{\text{up}} S_{\text{up}} i_{\text{down}} S_{\text{down}};S_{\text{ud}} M_{\text{ud}} \right\rangle \left|
S_lS_r;S_{lr}M_{lr}\right\rangle .  
\label{2_15}
\end{equation}
The reduced matrix elements of spin operators needed to build the Hamiltonian of the environment are relegated to Appendix [see Eqs.(\ref{A_2}-\ref{A_7})]. Note, that a number of the states (\ref{2_15}) is too much to avoid the truncation procedure (see Sec.III).

Matrix elements of the environment Hamiltonian 
$H_{\text{env}}=H_l+H_r+H_{\text{down}}+H_{\text{up}}+V_{\text{env}}$ are given as follows 
\[
\left\langle \left( i_{\text{up}}S_{\text{up}} i_{\text{down}} S_{\text{down}} \right) S_{\text{ud}},\left( S_lS_r\right)
S_{lr};S_{\text{env}}M_{\text{env}}|H_{\text{env}}|\left( i_{\text{up}}^{\prime
}S_{\text{up}}^{\prime }i_{\text{down}}^{\prime }S_{\text{down}}^{\prime }\right) S_{\text{ud}}^{\prime },\left(
S_l^{\prime }S_r^{\prime }\right) S_{lr}^{\prime };S_{\text{env}}^{\prime
}M_{\text{env}}^{\prime }\right\rangle 
\]
\[
=\left( E_{i_{\text{up}}S_{\text{up}}}+E_{i_{\text{down}} S_{\text{down}}}+E_{S_l}+E_{S_r}\right) \delta
_{i_{\text{up}},i_{\text{up}}^{\prime }}\delta _{S_{\text{up}},S_{\text{up}}^{\prime }}\delta _{i_{\text{down}},i_{\text{down}}^{\prime
}}\delta _{S_{\text{down}},S_{\text{down}}^{\prime }}\delta _{S_{\text{ud}},S_{\text{ud}}^{\prime }}\delta
_{S_l,S_l^{\prime }}\delta _{S_r,S_r^{\prime }}\delta
_{S_{lr},S_{lr}^{\prime }}\delta _{S_{\text{env}},S_{\text{env}}^{\prime
}}\delta _{M_{\text{env}},M_{\text{env}}^{\prime }} 
\]
\begin{equation}
+J_1\delta _{S_{\text{env}},S_{\text{env}}^{\prime }}(-1)^{S_{\text{env}%
}+S_{\text{ud}}^{\prime }+S_{lr}}\left\{ 
\begin{array}{ccc}
S_{\text{ud}} & S_{lr} & S_{\text{env}} \\ 
S_{lr}^{\prime } & S_{\text{ud}}^{\prime } & 1
\end{array}
\right\} \delta _{M_{\text{env}},M_{\text{env}}^{\prime }}  
\label{2_16}
\end{equation}
\[
\times \left\{
 \left\langle S_lS_r;S_{lr}\left\| S_{c1}\right\| S_l^{\prime
}S_r^{\prime };S_{lr}^{\prime }\right\rangle \left\langle
i_{\text{up}} S_{\text{up}} i_{\text{down}} S_{\text{down}};S_{\text{ud}}\left\| S_{c2}\right\| i_{\text{up}}^{\prime } S_{\text{up}}^{\prime
}i_{\text{down}}^{\prime }S_{\text{down}}^{\prime };S_{\text{ud}}^{\prime }\right\rangle 
\right.
\]
\[
+ \left\langle S_lS_r;S_{lr}\left\| S_{d1}\right\| S_l^{\prime
}S_r^{\prime };S_{lr}^{\prime }\right\rangle \left\langle
i_{\text{up}} S_{\text{up}} i_{\text{down}} S_{\text{down}};S_{\text{ud}}\left\| S_{d2}\right\| i_{\text{up}}^{\prime }S_{\text{up}}^{\prime
}i_{\text{down}}^{\prime }S_{\text{down}}^{\prime };S_{\text{ud}}^{\prime }\right\rangle 
\]
\[
+ \left\langle S_lS_r;S_{lr}\left\| S_{c5}\right\| S_l^{\prime
}S_r^{\prime };S_{lr}^{\prime }\right\rangle \left\langle
i_{\text{up}} S_{\text{up}} i_{\text{down}}S_{\text{down}};S_{\text{ud}}\left\| S_{c4}\right\| i_{\text{up}}^{\prime }S_{\text{up}}^{\prime
}i_{\text{down}}^{\prime }S_{\text{down}}^{\prime };S_{\text{ud}}^{\prime }\right\rangle 
\]
\[
\left.
+ \left\langle S_lS_r;S_{lr}\left\| S_{d5}\right\| S_l^{\prime
}S_r^{\prime };S_{lr}^{\prime }\right\rangle \left\langle
i_{\text{up}} S_{\text{up}} i_{\text{down}} S_{\text{down}};S_{\text{ud}}\left\| S_{d4}\right\| i_{\text{up}}^{\prime } S_{\text{up}}^{\prime
}i_{\text{down}}^{\prime }S_{\text{down}}^{\prime };S_{\text{ud}}^{\prime } \right \rangle  
\right\}. 
\]
The terms in $\{ \ldots \}$ include product of the reduced matrix
elements given by Eqs.(\ref{A_2},\ref{A_3}) for spins that enter into the left/right dimers
and by Eq.(\ref{A_6},\ref{A_7}) for the constituents of the fork-like parts.

After a finding of the environment 
eigenvalues $E_{i_{\text{env}}S_{\text{env}}}$ and eigenfunctions 
\begin{equation}
\left| i_{\text{env}}S_{\text{env}}M_{\text{env}}\right\rangle =\sum \beta
_{\left( i_{\text{up}}S_{\text{up}} i_{\text{down}} S_{\text{down}} \right) S_{\text{ud}},\left( S_lS_r\right) S_{lr}}^{i_{\text{%
env}}S_{\text{env}}}\left| \left( i_{\text{up}} S_{\text{up}} i_{\text{down}} S_{\text{down}} \right) S_{\text{ud}},\left(
S_lS_r\right) S_{lr};S_{\text{env}}M_{\text{env}}\right\rangle ,
\label{2_17}
\end{equation}
one calculate within the basis the reduced matrix elements for the
environment spins that directly interact with the central dimer see [Eq.(\ref{A_4})].

At the final step of the diagonalization procedure one build the basis of
the entire cluster 
\[
\left| i_{\text{env}}S_{\text{env}},S_c;SM  \right\rangle
=\sum\limits_{M_{\text{env}}M_{c}}
\left[ 
\begin{array}{ccc}
S_{\text{env}} & S_{c} & S \\ 
M_{\text{env}} & M_{c} & M
\end{array}
\right] 
\left| i_{\text{env}} S_{\text{env}} M_{\text{env}} \right\rangle  \left| S_cM_c\right\rangle , 
\]
and determine the matrix elements of the cluster Hamiltonian (\ref{2_8}) 
\[
\left\langle i_{\text{env}}S_{\text{env}},S_c;SM|H|i_{\text{env}}^{\prime
}S_{\text{env}}^{\prime },S_c^{\prime };S^{\prime }M^{\prime }\right\rangle
=\left( E_{i_{\text{env}}S_{\text{env}}}+E_{S_c}\right) \delta _{i_{\text{env%
}},i_{\text{env}}^{\prime }}\delta _{S_{\text{env}},S_{\text{env}}^{\prime
}}\delta _{S_c,S_c^{\prime }}\delta _{S,S^{\prime }}\delta _{M,M^{\prime }} 
\]
\[
+ J_1\left( -1\right) ^{S+S_{\text{env}}^{\prime }+S_c}\left\{ 
\begin{array}{ccc}
S_{\text{env}} & S_c & S \\ 
S_c^{\prime } & S_{\text{env}}^{\prime } & 1
\end{array}
\right\} \delta _{S,S^{\prime }}\delta _{M,M^{\prime }} 
\]
\begin{equation}
\times \left[ \left\langle S_c\left\| S(1)\right\| S_c^{\prime
}\right\rangle \sum\limits_{k=c2,c4}\left\langle i_{\text{env}}S_{\text{env}%
}\left\| S_k\right\| i_{\text{env}}^{\prime }S_{\text{env}}^{\prime
}\right\rangle +\left\langle S_c\left\| S(2)\right\| S_c^{\prime
}\right\rangle \sum\limits_{k=d2,d4}\left\langle i_{\text{env}}S_{\text{env}%
}\left\| S_k\right\| i_{\text{env}}^{\prime }S_{\text{env}}^{\prime
}\right\rangle \right] ,  \label{2_19}
\end{equation}
where the RMEs are previously derived [see Eqs.(\ref{2_10}-\ref{2_11}) and 
Eq.(\ref{A_4})]. Numerical diagonalization of the matrix (\ref{2_19}) yields the
target spectrum $E_{iS}$ and the eigenfunctions 
\begin{equation}
\left| iSM\right\rangle =\sum \gamma_{S_{\text{env}}M_{\text{env}%
},S_c}^{iS}\left| S_{\text{env}}M_{\text{env}},S_c;SM\right\rangle .
\label{2_17}
\end{equation}

\section{Truncation procedure}

The classification of eigenstates of parts used to gather the total cluster
according to irreducible representations of $SU(2)$-group enables to
organize a truncation procedure inside the sectors of Hilbert space that
arise at the consecutive steps of the algorithm. A possibility to carry out
calculations within a reduced basis is feature of the algorithm that relates
it with other renormalization group methods.

We hold the following strategy of the truncation procedure to build target
states that are obtained after combining two parts of the lattice. For a given spin-$S$ sector a certain amount of states having the lowest energy are kept. Thus every group of the $\left|iS\right\rangle$ states is presented in the reduced basis. 
We truncate the basis of two "fork"-like parts  before to combine them into a larger lattice segment. This is not the unique way, for example one can truncate the basis of the environment after combining the "fork"-like parts, but the former is easier to perform.  

 We tested  several realizations of the truncation procedure either by simply controlling a number of vectors retained in the reduced basis or by monitoring a genealogy of the target spin-$S$ state through the triangle rule, i.e. only states that contribute into the target state are took into account. The last approach gives an opportunity to keep more vectors in the basis due to an omitting of redundant states. Moreover, highest-spin cluster states, i.e. those with $S\geq 15$ in our problem, are treated exactly.   The size of truncated basis was chosen to be equal to either  64 or 121 for the scheme without an account of genealogy of the target state, and it varies from 12 till 352, being depending on the total spin $S$, for the "genealogical" scheme. 
 
 An accuracy of the truncation procedure is controlled by monitoring an energy of the lowest state within an each spin sector. The variation of this observable computed through the both schemes does  not usually exceed 1-2\% (a maximal discrepancy of order 6\% is reached only in the $S$-8 sector) that evidences a correctness of the constructed basis which exhibits almost no dependence on the used truncation procedure.  The results that we present below are obtained within the "genealogical" scheme.

Another feature of the algorithm is an addition of a central unit (one site
or dimer) with its environment at the final step. The procedure does not
depend on a structure of the environment and looks similar for any cluster.
However, the information about quantum numbers of the environment states enables  to simplify  calculations substantially at the stage of the algorithm. Indeed, for a given spin-$S$ sector of the Hilbert space of the
entire cluster one should pick out only those environment eigenfunctions
whose spins $S_u$ obey the rule 
\[
|S_u-S_c|\leq S\leq S_u+S_c. 
\]

Using of the truncation procedure results in the basises composed maximally from 4-5  thousand states.  To  control an  accuracy of the procedure  the  results obtained for the 18-site system are compared  with those for  10-site system.  The smaller cluster enables to handle a whole basis without any truncation.  The  10-site system is embedded into the bigger cluster and consists of the following parts: the central dimer $c3,d3$ and  the neighbor dimers  $b2,c2$, $b4,c4$, $d2,e2$ and $d4,e4$.  Apparently, a construction of the environment requires two consecutive steps (i) an addition of the dimers $b2,c2$ and  $b4,c4$  as well as $d2,e2$ and $d4,e4$ ones according to Eq.(\ref{2_13}) then followed by a calculation of reduced matrix elements according to Eq.(\ref{A_2},\ref{A_3}); (ii) a construction of the environment states from the upper and down parts built previously and a calculation of  RME of the environment spins that interact directly with the central dimer. The 
 entire cluster Hamiltonian is obtained through (\ref{2_19}). The biggest Hilbert space  dimension ($2025 \times 2025$) is reached in the $S$-2 sector. The numerical results for the supplementary cluster are listed in Table I for comparison. Note that one should compare energy values with the same magnetization per dimer (See Fig. \ref{cusp}).

\section{Energy spectrum and magnetization curve}

\begin{figure}[ptb]
\begin{center}
\includegraphics[width=85mm]{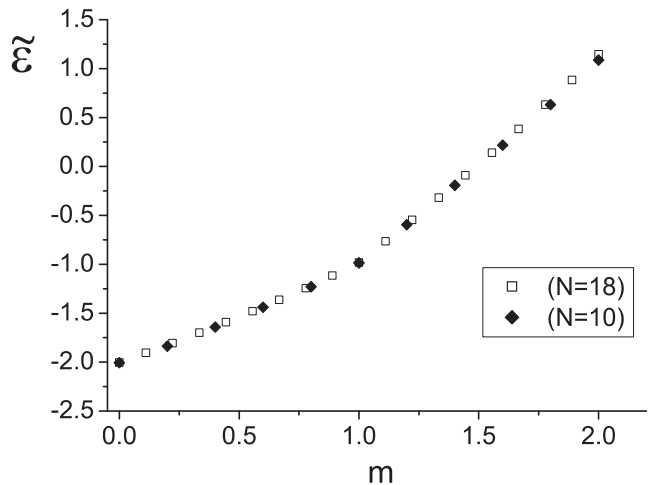}
\end{center}
\caption{Plot of the lowest energy per dimer $\tilde{\varepsilon}\,(m)$ vs $m$ for the $N=10$ and $N=18$ clusters. The cusp is seen at $m=1$.}
\label{cusp}
\end{figure}

The results of the energy spectrum calculation for two $N=10$ and $N=18$
clusters are listed in Table I, where we give minimal energy $E_{\min }$
within the each spin-$S$ sector along with the energy per dimer $\tilde{%
\varepsilon}\,=2E_{\min }/N$. The magnetization per dimer is determined by $%
m=2S/N$. The $N=10$ and $N=18$ dependencies $\tilde{\varepsilon}\,(m)$ are
shown together in Fig. \ref{cusp}. The points for both clusters lay on one
curve, i.e. finite-size effects may be ignored that is expected for the
regime of a small dimer-dimer interaction $J_1 \ll J_0$.

A remarkable feature of the curve is a cusp in the middle, i.e. at $m=1$.
The independent fitting of both parts jointed in the point by the quadratic
form $\varepsilon (m)=\varepsilon _2m^2+\varepsilon _1m+\varepsilon _0$
yields $\varepsilon _2=0.190\pm 0.018$, $\varepsilon _1=0.828\pm 0.019$, and 
$\varepsilon _0=-2.0073\pm 0.0040$ for the lower part of the curve ($0<m<1$)
together with $\varepsilon _2=0.200\pm 0.058$, $\varepsilon _1=1.4578\pm
0.018$, and $\varepsilon _0=-2.629\pm 0.014$ for the upper part ($1<m<2$).

\begin{figure}[ptb]
\begin{center}
\includegraphics[width=85mm]{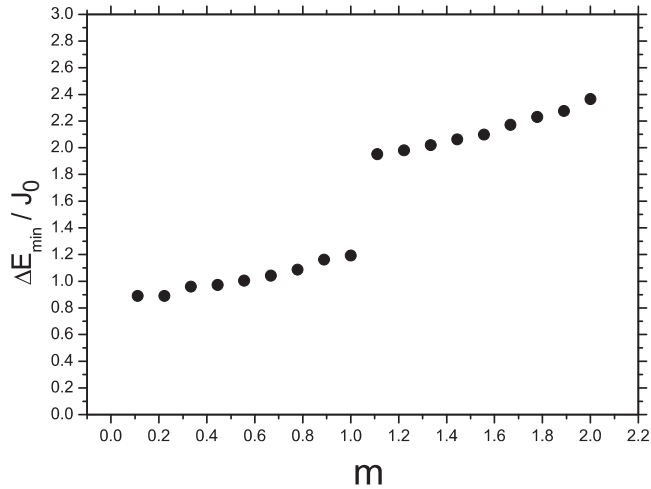}
\end{center}
\caption{Plot of the changes $\Delta E_{\min }$ versus the dimer magnetization $m$. A distinct jump is seen at $m=1$.}
\label{jump}
\end{figure}

On the base of data for N=18 case we build a dependence of jumps $E_{\min }$
when the total spin $S$ changes from $0$ till $18$, or the dimer
magnetization varies from $0$ till $2$ (Fig. \ref{jump}) One can see that the values of
the jumps are approximately $J_0$ for $S\leq 9$ and they are increased by a
factor of $2$ as $S\geq 10$. It means that the energy of the total system of
weakly interacting dimers will change with an increasing magnetic field due
to local excitations inside separate dimers. Indeed, for the single $S=1$
dimer the spectrum consists of a singlet, a triplet, and a quintuplet. The
energy difference between the singlet and triplet is $J_0$ while the
difference between the quintuplet and the triplet is $2J_0$ (see discussion
in the next Section).

\begin{table}
\caption{Numerical results of the lowest energy $E_{\min }$  and the energy $\tilde{\varepsilon}$ per dimer  in the spin-$S$ subspaces for $N=10$ and $N=18$ clusters.}
$
\begin{array}{|c|c|c|c|c|}
\hline
S & E_{\min }(N=10)/J_0 & \tilde{\varepsilon}\,(N=10) & E_{\min }(N=18)/J_0
& \tilde{\varepsilon}\,(N=18) \\ \hline
0 & -10.0334 & -2.0067 & -18.0336 & -2.0037 \\ 
1 & -9.1853 & -1.8371 & -17.1431 & -1.9048 \\ 
2 & -8.2123 & -1.6425 & -16.2529 & -1.8059 \\ 
3 & -7.1978 & -1.4396 & -15.2935 & -1.6993 \\ 
4 & -6.1430 & -1.2286 & -14.3205 & -1.5912 \\ 
5 & -4.9344 & -0.9869 & -13.3164 & -1.4796 \\ 
6 & -2.9787 & -0.5957 & -12.2745 & -1.3638 \\ 
7 & -0.9610 & -0.1922 & -11.1879 & -1.2431 \\ 
8 & 1.0849 & 0.2170 & -10.0260 & -1.1140 \\ 
9 & 3.1588 & 0.6318 & -8.8335 & -0.9815 \\ 
10 & 5.4418 & 1.0883 & -6.8807 & -0.7645 \\ 
11 &  &  & -4.8994 & -0.5444 \\ 
12 &  &  & -2.8795 & -0.3199 \\ 
13 &  &  & -0.8172 & -0.0908 \\ 
14 &  &  & 1.2815 & 0.1424 \\ 
15 &  &  & 3.4533 & 0.3837 \\ 
16 &  &  & 5.6844 & 0.6316 \\ 
17 &  &  & 7.960 & 0.8844 \\ 
18 &  &  & 10.3254 & 1.1473 \\ \hline
\end{array}
$
\label{energy}
\end{table}

\begin{figure}[ptb]
\begin{center}
\includegraphics[width=85mm]{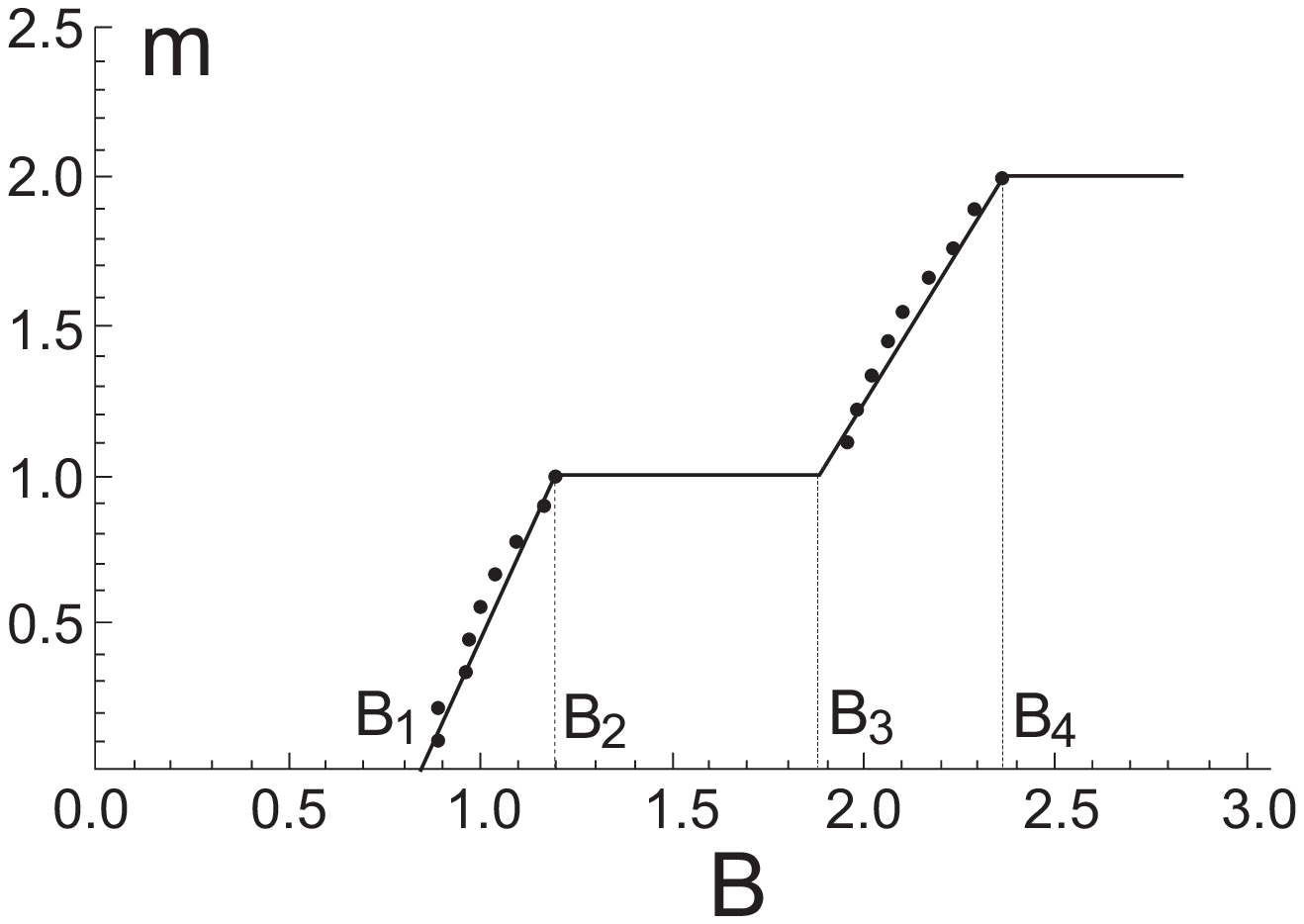}
\end{center}
\caption{Plot of $m$ versus $B$ obtained via $B=\varepsilon^{\prime} (m)$. The dots mark values found through the diagonalization algorithm.}
\label{plato}
\end{figure}

A standard way to consider a magnetization process at $T=0$ is to define $%
E_{\min }^{(S)}(N)$ as the lowest energy of the Hamiltonian (\ref{Hspin}) in
the spin-$S$ subspace for a finite system of $N$ elementary dimers. Applying
a magnetic field $B$ leads to a Zeeman splitting of the energy levels $%
E_{\min }^{(S)}(B)=E_{\min }^{(S)}-SB$, and therefore level crossing occurs
at values $B_S=E_{\min }^{(S+1)}(B)-E_{\min }^{(S)}(B)$ when increasing the
field. These level crossings correspond to jumps in the magnetization at
zero temperature of the value $1/N$, until the fully polarized state with
the magnetization per dimer $m_{\text{sat}}=2N/N=2$ is reached at the value
of the magnetic field $B_{\text{sat}}=E_{\min }^{(2N)}(B)-E_{\min
}^{(2N-1)}(B)$. The calculation performed for $N/2=9$ dimers yields the magnetization 
points presented in Fig. \ref{plato} and demonstrates an appearance of the ground state
plateau as well as the plateau at one-half of the saturation value.

To guarantee a validity of the magnetization curve we use an approach
developed by Sakai and Tahakashi\cite{Takahashi} to recover the $m(B)$
dependence in the thermodynamical limit. In this case the condition for the
crossover fields transforms into $B=\varepsilon ^{\prime }(m)$, where $%
\varepsilon $ is the energy per dimer. The plateau boundaries are determined
by the derivatives in the special points: (i) $B_1=\varepsilon ^{\prime
}(+0) $ is related with the end of the ground state plateau; (ii) $%
B_2=\varepsilon ^{\prime }(1-0)$ and $B_3=\varepsilon ^{\prime }(1+0)$
correspond to the beginning and the end of the intermediate plateau,
respectively; (iii) $B_4=\varepsilon ^{\prime }(2-0)$ marks an emergence of
the saturation magnetization.

A treatment of the energy spectrum results in the linear dependences
relevant for sectors between the plateaus 
\begin{equation}
\left\{ 
\begin{array}{c}
\varepsilon ^{\prime }(m)=0.83+0.38\,m,\;0<m<1, \\ 
\varepsilon ^{\prime }(m)=1.46+0.40\,m,\;1<m<2,
\end{array}
\right.  \label{Btlm}
\end{equation}
that produces immediately $B_1=0.83\,J_0$, $B_2=1.21\,J_0$, $B_3=1.86\,J_0$,
and $B_4=2.26\,J_0$. The values normalized to the saturation field $B_{\text{%
sat}}$ are listed in the Table II and exhibit a reasonable agreement with the
experimental data for F$_2$PNNNO system. A comparison of the finite cluster
calculations with those of the thermodynamical limit (\ref{Btlm}) is given
in Fig. \ref{plato}. It is seen that both methods produce the close results.

Note that the method we used for numerical calculations is intrinsically two-dimensional one whereas the previous numerical study of the system\cite{Hosokoshi} dealt with the  cluster embedded into a chain. The regions between the plateaus of the magnetization curve  exhibit a behavior  closer to linear one instead of the S-shape forms early obtained.

\begin{table}
\caption{Values of the magnetic field  special points compared with the experimental data.}
$
\begin{array}{|c|c|c|c|c|}
\hline
B_i/B_{\text{sat}} & i=1 & i=2 & i=3 & i=4 \\ \hline
\text{Theory} & 0.37 & 0.53 & 0.82 & 1 \\ \hline
\text{Experiment\cite{Hosokoshi}} & 0.33 & 0.53 & 0.89 & 1 \\ \hline
\end{array}
$
\label{B_values}
\end{table}

\section{Semi-hard core boson model}

Let us introduce the boson picture on the base of data presented in Fig. \ref{jump}.
For $J_1 \ll J_0$ the low energy subspace of spin Hamiltonian (\ref{Hspin})
consists of the singlet, the $S^z=1$ component of the triplet, and the $%
S^z=2 $ component of the quintuplet. It is convenient to identify the
triplet state with a presence of a bosonic particle (triplon), the
quintiplet state as a pair of bosons (quintuplon), and the singlet state as
an absence of bosons. Then the boson model is formulated via the semi-hard
core bosonic operators $g_i$ and $g_i^{\dagger }$ with the extended Pauli's
exclusion principle $g_i^{\dagger \,3}=0$, i.e. more then two bosons per site are
forbidden. Note that the principle may be realized via parafermion language but the description requires a transmutation of statistics that complicates calculations in a 2D case (see Appendix B).  The algebra of the operators are $\left[ g_i,g_i\right] =\left[
g_i^{\dagger },g_i^{\dagger }\right] =0$, and $\left[ g_i,g_i^{\dagger
}\right] =\delta _{ij}\left( 1-F_i\right) $, where $F_i=\left( 3/2\right)
n_i\left( n_i-1\right) $ is the deformation of the canonical boson algebra, $%
n_i=g_i^{\dagger }g_i$ is the number operator.\cite{Batista}

The boson Hamiltonian in terms of these operators is written as 
\begin{equation}
H=\frac 12\sum_{\left\langle ij\right\rangle }\left( g_i^{\dagger
}g_j+g_j^{\dagger }g_i\right) \left( h_1+h_2+h_3\right) -\mu \sum_in_i+\frac 
U2\sum_in_i\left( n_i+1\right) +V\sum_{\left\langle ij\right\rangle }\left(
n_i-1\right) \left( n_j-1\right) ,  \label{Hbos}
\end{equation}
where the hopping terms 
\[
h_1=t_1\left( n_{ij}-2\right) \left( n_{ij}-3\right) ,\;h_2=2t_2\left(
n_{ij}-1\right) \left( 3-n_{ij}\right) ,\;h_1=t_3\left( n_{ij}-1\right)
\left( n_{ij}-2\right) 
\]
depend on a number of particles $n_{ij}=n_i+n_j$ on the bonds $i,j$.

\begin{figure}[ptb]
\begin{center}
\includegraphics[width=85mm]{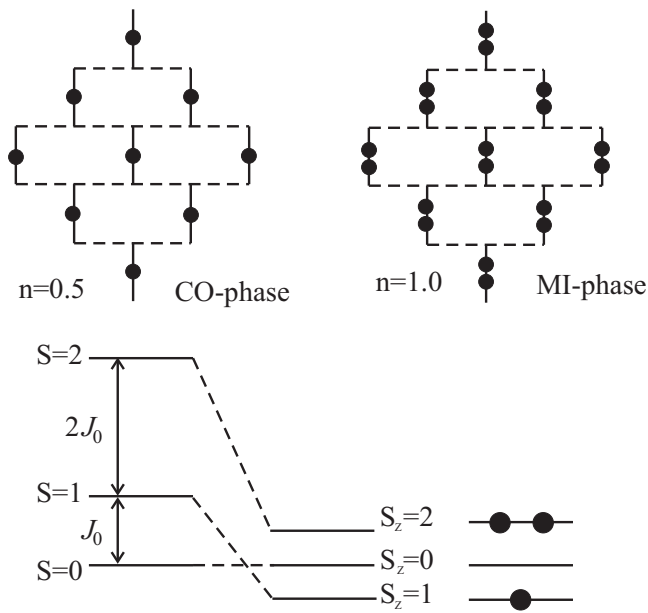}
\end{center}
\caption{The low energy subspace of the single dimer spectrum in the presence of a magnetic field. Boson superlattice patterns corresponding to the charge-ordered and Mott insulating phases are shown above.}
\label{onsite}
\end{figure}

The map between the bosonic (\ref{Hbos}) and the spin Hamiltonian (\ref
{Hspin}) is reached through the representation\cite{Sengupta} (see Fig. \ref{onsite})

\[
n_i=S_i^z=S_{i1}^z+S_{i2}^z, 
\]
where $(1,2)$ marks two spins on each dimer, and 
\[
g_i^{\dagger }=\frac 1{\sqrt{2}}\left( S_{i2}^{\dagger }-S_{i1}^{\dagger
}\right) \left[ \frac{\sqrt{3}}{2\sqrt{2}}+\left( 1-\frac{\sqrt{3}}{2\sqrt{2}%
}\right) S_i^z\right] . 
\]
This establishes the relationship between the spin and the bosonic
parameters $U=J_0$, $V=J_1/2$, $\mu =B-4J_1$, and $t_i=-\frac{8\sqrt{2}}{3%
\sqrt{3}}a^iJ_1$, where $a=\frac{\sqrt{3}}{2\sqrt{2}}$ $(i=1,2,3)$. Thus,
the bosonic model includes the strong on-site boson repulsion $U$ as well as
the noticeable repulsive intersite interaction $V$. The magnetic field $B$
plays the role of the chemical potential $\mu $.

The boson Hamiltonian (\ref{Hbos}) constitutes low-energy effective model of the spin Hamiltonian (\ref{Hspin}) that appears from restricting $H_S$ to the subspace of the semi-hard core bosonic operators. The map is valid in the limit $J_1 \ll J_0$, or in the boson language $t_i/U$, $V/U \ll 1$, when the main physics is governed by a competition between  the one-site repulsion and the chemical potential.

The quantum phase diagram of the boson Hamiltonian (\ref{Hbos})  was built in Ref.\cite{Sengupta} by using the stochastic
series expansion quantum Monte Carlo method (see there Fig.4). It has been found that a Bose condensate fraction appears in the regions of the chemical potential (magnetic field) between the platos of the $g$-particles density (magnetization curve). In contrast, the charge density wave (Ising-like charge order (CO) phase) forms around the intermediate plato. There are regions, where supersolid phase, a mixing of the charge order and the Bose-supefluid (BS), emerges. According to the study the magnetization curve shown in Fig. 4 can be interpreted as a tuning of boson density
by the applied magnetic field. At small chemical potential the empty states
has the lowest energy, when all dimers in the singlet state (boson vacuum).
For $B>B_1$ a finite density of bosons (triplons) emerges in the ground
state and contributes into a BS phase. The triplon
excitations are mobile due to weak interdimer coupling. The density
(magnetization) increases monotonically as a function of magnetic field
until $B_2$, where a transition to the  CO-phase comes up. This
corresponds to the boson concentration $n=0.5$, when the triplons crystallize
in a superstructure pattern (Fig. \ref{onsite}). The fractional plateau requires strong
boson interactions in comparison to the kinetic energy.  At $B>B_3$ the
filling increases monotonically in the resulting BS phase (quintiplon
condensation) till the ground state transforms into a Mott insulating (MI)
phase with two bosons per dimer at $B>B_4$. The boson concentration in the
MI phase $n=1$. The reasonings are easily reproduced  if to analyze the boson Hamiltonian (\ref{Hbos}) by neglecting the intersite terms.

\section{Conclusions}

Quantum dimer antiferromagnetic systems is a nice testing area to study BEC
of interacting particles. Along with the ultracold atomic gases in optical
lattices\cite{Grenier,Bloch} they offer an opportunity to observe
transitions predicted by lattice boson models. In many problems the
boson picture is more physically transparent than the original spin
language. On the base of the analysis of the finite cluster energy spectrum
for the two-dimensional spin-1 organic antiferromagnet F$_2$PNNNO with the
dimerized structure we prove a relevance of the model of semi-hard core
bosons with pronounced on-site and inter-site repulsions for the
low-dimensional spin system. The unusual magnetization curve observed in F$%
_2 $PNNNO is nothing but a manifestation of fine-tune of density of the
bosons by the applied magnetic field, when the low-density Bose-superfluid,
charge ordering with one boson per a dimer, and high-density Bose-supefluid
phases change subsequently each other with an increasing of the field.

\begin{acknowledgments}
We would like to thank T. Sakai,  J. Kishine and N.V. Baranov for
discussions. V.E.S. would like to acknowledge the support of the U.S. Civilian Research \& Development Foundation (CRDF) and Ministery of Education and Science of Russian Federation (MinES) under "Basic Research and Higher Education" (BRHE) program.
\end{acknowledgments}

\appendix

\section{}

The reduced matrix elements for spins on the $c2$ and $c4$ sites computed in the basis of  eigenfunctions of the Hamiltonian  $H_{\text{down}}$ are given by the 141 $\times$ 141 matrix
$$
\left\langle i_{\text{down}}S_{\text{down}}\left\| S_{c2(c4)}\right\| i_{\text{down}}^{\prime }S_{\text{down}}^{^{\prime
}}\right\rangle =\sum\limits_{S_2S_3S_4S_{24}}\sum\limits_{S_2^{\prime
}S_3^{\prime }S_4^{\prime }S_{24}^{\prime }}\alpha _{\left( S_2S_4\right)
S_{24},S_3}^{i_{\text{down}}S_{\text{down}}}\alpha _{\left( S_2^{\prime }S_4^{\prime }\right)
S_{24}^{\prime },S_3^{\prime }}^{i_{\text{down}}^{\prime }S_{\text{down}}^{\prime }} 
$$
\begin{equation}
\times \left\langle \left( S_2S_4\right) S_{24},S_3;S_{\text{down}}\left\|
S_{c2(c4)}\right\| \left( S_2^{\prime }S_4^{\prime }\right) S_{24}^{\prime
},S_3^{\prime };S_{\text{down}}^{\prime }\right\rangle. 
 \label{A_1}
\end{equation} 

The reduced matrix elements that enter into the expression are calculated
according to the rules 
\[
\left\langle \left( S_2S_4\right) S_{24},S_3;S_{\text{down}}\left\| S_{c2}\right\|
\left( S_2^{\prime }S_4^{\prime }\right) S_{24}^{\prime },S_3^{\prime
};S_{\text{down}}^{\prime }\right\rangle =(-1)^{S_2+S_4+S_3+S_{24}+S_{24}^{\prime
}+S_{\text{down}}^{\prime }}\left[ S_{24},S_{24}^{\prime },S_{\text{down}},S_{\text{down}}^{\prime }\right]
^{1/2} 
\]
\[
\times \left\{ 
\begin{array}{ccc}
S_{24} & 1 & S_{24}^{\prime } \\ 
S_2^{\prime } & S_4 & S_2
\end{array}
\right\} \left\{ 
\begin{array}{ccc}
S_{\text{down}} & 1 & S_{\text{down}}^{\prime } \\ 
S_{24}^{\prime } & S_3 & S_{24}
\end{array}
\right\} \left\langle 11;S_2\left\| S(2) \right\| 11;S_2^{\prime
}\right\rangle \delta _{S_4S_4^{\prime }}\delta _{S_3S_3^{\prime }}, 
\]

\[
\left\langle \left( S_2S_4\right) S_{24},S_3;S_{\text{down}}\left\| S_{c4}\right\|
\left( S_2^{\prime }S_4^{\prime }\right) S_{24}^{\prime },S_3^{\prime
};S_{\text{down}}^{\prime }\right\rangle =(-1)^{S_2+S_4^{\prime
}+S_3+2S_{24}+S_{\text{down}}^{\prime }}\left[ S_{24},S_{24}^{\prime },S_{\text{down}},S_{\text{down}}^{\prime
}\right] ^{1/2} 
\]
\[
\times \left\{ 
\begin{array}{ccc}
S_{24} & 1 & S_{24}^{\prime } \\ 
S_4^{\prime } & S_2 & S_4
\end{array}
\right\} \left\{ 
\begin{array}{ccc}
S_{\text{down}} & 1 & S_{\text{down}}^{\prime } \\ 
S_{24}^{\prime } & S_3 & S_{24}
\end{array}
\right\} \left\langle 11;S_4\left\| S(2)\right\| 11;S_4^{\prime
}\right\rangle \delta _{S_2S_2^{\prime }}\delta _{S_3S_3^{\prime }}, 
\]
where $[ S ]=  (2S+1)$. 

The reduced matrix elements for spins on the sites $c1$ ($d1$) are given by the $19 \times 19 $ matrix  built in the basis of functions constructed from the "left" and the "right" dimers Eq.(\ref{2_13})
\begin{equation} 
\left\langle S_l S_r ;S_{lr}\left\|S_{c1(d1)}\right\|S_l^{\prime }S_r^{\prime };S_{lr}^{\prime }\right\rangle=\sqrt{(2S_{lr}+1)(2S_{lr}^{\prime}+1)}(-1)^{1+S_l+S_r+S_{lr}^{\prime}}
\left\{ 
\begin{array}{ccc}
S_{lr} & 1 & S_{lr}^{\prime } \\ 
S_l^{\prime } & S_r & S_l
\end{array}
\right\} \left\langle 11;S_l\left\| S(1(2))\right\| 11; S_l^{\prime
}\right\rangle \delta _{S_r S_r^{\prime }}. 
\label{A_2}
\end{equation} 
The RME for spins on the $c5$ ($d5$) sites amount to
\begin{equation}
\left\langle S_l S_r ;S_{lr}\left\|S_{c5(d5)}\right\|S_l^{\prime }S_r^{\prime };S_{lr}^{\prime }\right\rangle=\sqrt{(2S_{lr}+1)(2S_{lr}^{\prime}+1)}(-1)^{1+S_l+S_r^{\prime}+S_{lr}}
\left\{ 
\begin{array}{ccc}
S_{lr} & 1 & S_{lr}^{\prime } \\ 
S_r^{\prime } & S_l & S_r
\end{array}
\right\} \left\langle 11;S_r\left\| S(1(2))\right\| 11;S_r^{\prime
}\right\rangle \delta _{S_lS_l^{\prime }}.
\label{A_3}
\end{equation} 
The reduced matrix elements of spin operators on sites $c2(d2),c4(d4)$ calculated on the eigenfuctions of the upper and down parts form the $73789 \times 73789$ matrices. 
$$
\left\langle
i_{\text{up}}S_{\text{up}}i_{\text{down}}S_{\text{down}};S_{\text{ud}}\left\| S_{c2(c4)}\right\| i_{\text{up}}^{\prime }S_{\text{up}}^{\prime
}i_{\text{down}}^{\prime }S_{\text{down}}^{\prime };S_{\text{ud}}^{\prime }\right\rangle  =
\sqrt{(2S_{\text{ud}}+1)(2S_{\text{ud}}^{\prime}+1)}(-1)^{1+S_{\text{up}}+S_{\text{down}}^{\prime}+S_{\text{ud}}}
$$
\begin{equation}
\left\{ 
\begin{array}{ccc}
S_{\text{ud}} & 1 & S_{\text{ud}}^{\prime } \\ 
S_{\text{down}}^{\prime } & S_{\text{up}} & S_{\text{down}}
\end{array}
\right\} 
\left\langle i_{\text{down}}S_{\text{down}}\left\| S_{c2(c4)}\right\| i_{\text{down}}^{\prime }S_{\text{down}}^{^{\prime
}}\right\rangle
\delta _{i_{\text{up}}i_{\text{up}}^{\prime }} \delta _{S_{\text{up}}S_{\text{up}}^{\prime }}.
\label{A_6}
\end{equation} 

$$
\left\langle
i_{\text{up}}S_{\text{up}}i_{\text{down}}S_{\text{down}};S_{\text{ud}} \left\| S_{d2(d4)} \right\| i_{\text{up}}^{\prime }S_{\text{up}}^{\prime
} i_{\text{down}}^{\prime }S_{\text{down}}^{\prime };S_{\text{ud}}^{\prime }\right\rangle  =
\sqrt{(2S_{\text{ud}}+1)(2S_{\text{ud}}^{\prime}+1)}(-1)^{1+S_{\text{up}}+S_{\text{down}}+S_{\text{ud}}^{\prime}}
$$
\begin{equation}
\left\{ 
\begin{array}{ccc}
S_{\text{ud}} & 1 & S_{\text{ud}}^{\prime } \\ 
S_{\text{up}}^{\prime } & S_{\text{down}} & S_{\text{up}}
\end{array}
\right\} 
\left\langle i_{\text{up}}S_{\text{up}}\left\| S_{d2(d4)}\right\| i_{\text{up}}^{\prime }S_{\text{up}}^{^{\prime
}}\right\rangle
\delta _{i_{\text{down}}i_{\text{down}}^{\prime }} \delta _{S_{\text{down}}S_{\text{down}}^{\prime }}.
\label{A_7}
\end{equation}

The reduced matrix elements of spin operators on sites $c2(d2),c4(d4)$ calculated on the eigenfuctions of the environment. The dimension of this matrices determines by dimension of truncated basis of environment
\[
\left\langle i_{\text{env}}S_{\text{env}}\left\| S_k\right\| i_{\text{env}%
}^{\prime }S_{\text{env}}^{\prime }\right\rangle =\sum \beta _{\left(
i_{\text{up}}S_{\text{up}}i_{\text{down}}S_{\text{down}}\right) S_{\text{ud}},\left( S_lS_r\right) S_{lr}}^{i_{\text{env}}S_{%
\text{env}}}\beta _{\left( i_{\text{up}}^{\prime }S_{\text{up}}^{\prime }i_{\text{down}}^{\prime
}S_{\text{down}}^{\prime }\right) S_{\text{ud}}^{\prime },\left( S_l^{\prime }S_r^{\prime
}\right) S_{lr}^{\prime }}^{i_{\text{env}}^{\prime }S_{\text{env}}^{\prime
}} 
\]
\begin{equation}
\times \left\langle \left( i_{\text{up}}S_{\text{up}}i_{\text{down}}S_{\text{down}}\right) S_{\text{ud}},\left( S_lS_r\right)
S_{lr};S_{\text{env}}\left\| S_k\right\| \left( i_{\text{up}}^{\prime }S_{\text{up}}^{\prime
}i_{\text{down}}^{\prime }S_{\text{down}}^{\prime }\right) S_{\text{ud}}^{\prime },\left( S_l^{\prime
}S_r^{\prime }\right) S_{lr}^{\prime };S_{\text{env}}^{\prime }\right\rangle
,  \label{A_4}
\end{equation}
where $k=c2(d2),c4(d4)$ and
$$
\left\langle \left( i_{\text{up}}S_{\text{up}}i_{\text{down}}S_{\text{down}}\right) S_{\text{ud}},\left( S_lS_r\right)
S_{lr};S_{\text{env}} \left\| S_k \right\| \left( i_{\text{up}}^{\prime }S_{\text{up}}^{\prime
}i_{\text{down}}^{\prime }S_{\text{down}}^{\prime }\right) S_{\text{ud}}^{\prime },\left( S_l^{\prime
}S_r^{\prime }\right) S_{lr}^{\prime };S_{\text{env}}^{\prime }\right\rangle=
\sqrt{(2S_{\text{env}}+1)(2S_{\text{env}}^{\prime}+1)}
$$
\begin{equation}
\times (-1)^{1+S_{\text{ud}}+S_{lr}+S_{\text{env}}^{\prime}}
\left\{ 
\begin{array}{ccc}
S_{\text{env}} & 1 & S_{\text{env}}^{\prime } \\ 
S_{\text{ud}}^{\prime } & S_{lr} & S_{\text{ud}}
\end{array}
\right\} \left\langle
i_{\text{up}}S_{\text{up}}i_{\text{down}}S_{\text{down}};S_{\text{ud}}\left\| S_k\right\| i_{\text{up}}^{\prime }S_{\text{up}}^{\prime
}i_{\text{down}}^{\prime }S_{\text{down}}^{\prime };S_{\text{ud}}^{\prime }\right\rangle  \delta _{S_lS_l^{\prime }}\delta _{S_rS_r^{\prime }}\delta _{S_{lr}S_{lr}^{\prime }}.
  \label{A_5}
\end{equation} 

\section{}

Quantum statistics is based on two principles, the first is the exchange statistics, when  a permutation of two identical particles causes an appearance of  a phase factor in the total wave function, and the second is the exclusion  statistics, which reflects an ability to accommodate $p$ particles in the same single-particle quantum state. Whereas the first concept depends on the space dimensionality of the system, the second one does not.\cite{Haldane1991}   

The exclusion statistics algebra obeying the generalized Pauli exclusion principle can be formulated in terms of the bond $g$ operators that has been used in the main text. Another variant of the exclusion statistics can be realized, for example, via Green's parafermion statistics.\cite{Green1953,Green1972} According to common formalism based on Burnside's theorem of the group theory (see Ref.\cite{Batista} for details) the both algebraic approaches are related with each other.

Indeed, let us introduce two modes ($\alpha=1,2$) for the each $i$-th bond
\be
\left\{
\begin{array}{c}
\left\{ d^{\alpha}_i, d^{\alpha}_j \right\} = \left\{ \left(d^{\alpha}_i\right)^{\dagger}, \left(d^{\alpha}_j\right)^{\dagger} \right\} = 0,\\
\left\{ d^{\alpha}_i, \left(d^{\alpha}_j\right)^{\dagger} \right\} = \delta_{ij}
\end{array}
\right.
\en 
with the condition $d^{\alpha}_j | \text{vacuum} \rangle = 0$. For $\alpha \not= \beta$ the modes satisfy non-standard relations 
\be
\left\{
\begin{array}{c}
\left[ d^{\alpha}_i, d^{\beta}_j \right] = \left[ \left(d^{\alpha}_i\right)^{\dagger}, \left(d^{\beta}_j\right)^{\dagger} \right] = 0,\\
\left[ d^{\alpha}_i, \left(d^{\beta}_j\right)^{\dagger} \right] = 0.
\end{array}
\right.
\en 
Parafermion creation and annihilation operators are determined as 
\be
  d_j^{\dagger} = \left(d^{1}_j\right)^{\dagger}+\left(d^{2}_j\right)^{\dagger}, \qquad   d_j = d^{1}_j +d^{2}_j.
\en
They satisfy the commutation relations
\be
\left\{
\begin{array}{c}
\left[ \left[ d^{\dagger}_i, d_j \right], d_l \right]  = -2 \delta_{il} d_j, \\
\left[ \left[ d_i, d_j \right], d_l \right]  = 0.
\end{array}
\right.
\en 
The parafermion number operator $n^d_j = \left(d^{1}_j\right)^{\dagger} d^{1}_j + \left(d^{2}_j\right)^{\dagger} d^{2}_j$ can be written as
\be
n^d_j = \frac12 \left( \left[ d^{\dagger}_i, d_j \right] + 2 \right),
\en
and obeys the commutation rule 
\be
\left[n^d_j, d^{\dagger}_j \right] = \delta_{ij}  d^{\dagger}_j.
\en
From the property $\left( n^{\alpha}_j \right)^2 = n^{\alpha}_j$ it follows that $n^d_j$ varies from 0 to $2$. Moreover, 
\be
\left( d^{\dagger}_j \right)^2 =  2 \left(d^1_i\right)^{\dagger} \left(d^2_i\right)^{\dagger},
\en
that means $\left( d^{\dagger}_j \right)^3=0$. Therefore, the parafermion representation provides the extended Pauli exclusion principle.

To  establish a connection between the bond $g$-algebra and the parafermion statistics, we note that the local  Hilbert space related with a bond has the  dimension $D=3$. Therefore, one can map the $g$-particles onto the algebra of $S$-1 operators
$$
S^{+}_i = \sqrt{2} g^{\dagger}_i \left[ 1+ \left(\frac{1}{\sqrt{2}} -1 \right) n^g_i \right],
$$
\be
S^{-}_i = \sqrt{2} \left[ 1+ \left(\frac{1}{\sqrt{2}} -1 \right) n^g_i \right] g_i,
\label{Sviag}
\en
$$
S^z_i = n^g_i -1.
$$

These spin operators are connected with two-flavour hard-core bosons via the generalization of the Jordan-Wigner transformation\cite{Batista2001,Sengupta2007} 
\be
\begin{array}{c}
S^{+}_i = \sqrt{2} \left( b^{\dagger}_{i1} + b_{i2} \right),\\
S^{-}_i = \sqrt{2} \left( b_{i1} + b^{\dagger}_{i2} \right),\\
S^z_i = b^{\dagger}_{i1} b_{i1} - b^{\dagger}_{i2} b_{i2}
\end{array}
\label{Batista}
\en
with the imposed constraint $b^{\dagger}_{i1}b_{i1} + b^{\dagger}_{i2}b_{i2} = 1$, and the spin state $S^z=0$ is taken as a vacuum. The commutation relations for the hard bosons are
$$
\left[ b_{i\alpha},b_{i\beta} \right] = \left[ b^{\dagger}_{i\alpha},b^{\dagger}_{i\beta} \right] = 0,  
$$
\be
\left[ b_{i\alpha},b^{\dagger}_{i\beta} \right] = \delta_{ij} \delta_{\alpha \beta} \left( 1 - n^b_{i\alpha} \right), \quad \left[ n^b_{i\alpha}, b^{\dagger}_{j\beta} \right] = \delta_{ij} \delta_{\alpha \beta} b^{\dagger}_{i\alpha}, 
\en
where $n^b_{i\alpha}=b^{\dagger}_{i\alpha} b_{i\alpha}$ ($\alpha=1,2$) is the  number operator for the hard bosons.

A transition from the hard-core bosons to the parafermions is related with a transmutation of statistics. In two-dimensional case the change of statistics is based on  a generalization of the conventional Jordan-Wigner transformation.\cite{Fradkin1989,Wang1991} In the following, for simplicity, we illustrate the connection on an example of dimerized one-dimensional S-1 chain.

The parafermion modes are converted into the canonical two-flavour canonical fermions $c_{i \alpha}$ ($\alpha=1,2$) determined on the $i$-th bond of the chain through the {\it partial} non-local transmutators
\be
\left(d^1_i\right)^{\dagger} = c^{\dagger}_{i1} \exp \left[ i \pi \sum_{j<i} n_{j2} \right], \quad 
\left(d^2_i\right)^{\dagger} = c^{\dagger}_{i2} \exp \left[ i \pi \left( \sum_{j<i} n_{j1} + n_{i1} \right) \right],
\label{dviac}
\en
where 
\be
\left\{ c_{i\alpha},c_{j\beta} \right\} = \left\{ c^{\dagger}_{i\alpha},c^{\dagger}_{j\beta} \right\} = 0, \quad 
\left\{ c_{i\alpha},c^{\dagger}_{j\beta} \right\} = \delta_{ij} \delta_{\alpha \beta},
\en
and $n_{i\alpha}=c^{\dagger}_{i\alpha} c_{i\alpha}$ is the  number operator for the fermions.

A map between the hard-core bosons and the two-flavour fermions is established by the {\it total} non-local transmutators
$$
b^{\dagger}_{i1} = c^{\dagger}_{i1}  \exp \left[ i \pi \sum_{j<i} \left( c^{\dagger}_{j1}c_{j1} + c^{\dagger}_{j2}c_{j2} \right) \right],
$$
\be
b^{\dagger}_{i2} = c^{\dagger}_{i2}  \exp \left[ i \pi \sum_{j<i} \left( c^{\dagger}_{j1}c_{j1} + c^{\dagger}_{j2}c_{j2} \right) \right] e^{i \pi c^{\dagger}_{j1}c_{j1}}.
\label{hbviac}
\en
The relations (\ref{Sviag},\ref{Batista},\ref{dviac},\ref{hbviac}) provide a map between the parafermions and the $g$-particles.

\end{document}